\documentclass[11pt,draftcls,onecolumn,twoside]{IEEEtran} % draft format for submission

\ifCLASSINFOpdf
\else
\fi

\usepackage{graphicx}      % include this line if your document contains figures
\usepackage{amssymb,amsmath,graphicx,multicol,epstopdf}

\DeclareMathOperator*\argmin{arg\,min}

\DeclareMathOperator*{\maximize}{maximize}

\newcommand{\R}{{\mathbb R}}
\newcommand{\inn}[1]{\langle #1 \rangle}

\newtheorem{Remark}{Remark}[section]
\newtheorem{Example}{Example}[section]

\newtheorem{Proposition}{Proposition}[section]

\newcommand{\mean}[1]{\mathbb{E}(#1)}
\newcommand{\var}[1]{\mathbb{V}\left(#1\right)}

\makeatletter
\renewcommand{\maketag@@@}[1]{\hbox{\m@th\normalsize\normalfont#1}}
\makeatother

\begin{document}
%
% paper title
% can use linebreaks \\ within to get better formatting as desired
% Do not put math or special symbols in the title.
\title{Continuous-time DC kernel ---\\ a stable generalized first-order spline kernel }
%
%
% author names and IEEE memberships
% note positions of commas and nonbreaking spaces ( ~ ) LaTeX will not break
% a structure at a ~ so this keeps an author's name from being broken across
% two lines.
% use \thanks{} to gain access to the first footnote area
% a separate \thanks must be used for each paragraph as LaTeX2e's \thanks
% was not built to handle multiple paragraphs
%

\author{Tianshi Chen,~\IEEEmembership{Member,~IEEE,}
%        Alessandro Chiuso,~\IEEEmembership{Senior Member,~IEEE,}
%        Gianluigi Pillonetto,~\IEEEmembership{Member,~IEEE,}
\thanks{T. Chen is with the School of Science and Engineering, The Chinese University of Hong Kong, Shenzhen, 518172, Shenzhen, China. \texttt{e-mail: tschen@cuhk.edu.cn}. He is also holding a visiting position at the Department of Electrical Engineering, Link\"{o}ping University, Link\"{o}ping, Sweden.
%Gianluigi Pillonetto is with
%the Department of Information Engineering, University of Padova.
%\texttt{e-mail: giapi@dei.unipd.it}.
}

\thanks{This work was supported by the Thousand Youth Talents Plan funded by the central
			government of China, the general project funded by NSFC under contract No. 61773329, the Shenzhen Research Projects Ji-20170189 and Ji-20160207 funded by Shenzhen Science and Technology Innovation Council, the
			President's grant under contract No. PF. 01.000249 and the start-up
			grant under contract No. 2014.0003.23 funded by the Chinese
			University of Hong Kong, Shenzhen, as well as by a research grant
			for junior researchers under contract No. 2014-5894, funded by
			Swedish Research Council.
}}

% note the % following the last \IEEEmembership and also \thanks -
% these prevent an unwanted space from occurring between the last author name
% and the end of the author line. i.e., if you had this:
%
% \author{....lastname \thanks{...} \thanks{...} }
%                     ^------------^------------^----Do not want these spaces!
%
% a space would be appended to the last name and could cause every name on that
% line to be shifted left slightly. This is one of those "LaTeX things". For
% instance, "\textbf{A} \textbf{B}" will typeset as "A B" not "AB". To get
% "AB" then you have to do: "\textbf{A}\textbf{B}"
% \thanks is no different in this regard, so shield the last } of each \thanks
% that ends a line with a % and do not let a space in before the next \thanks.
% Spaces after \IEEEmembership other than the last one are OK (and needed) as
% you are supposed to have spaces between the names. For what it is worth,
% this is a minor point as most people would not even notice if the said evil
% space somehow managed to creep in.

% The paper headers
\markboth{IEEE Transactions on Automatic Control,~Vol.~XX, No.~X, X~2017}%
{Shell \MakeLowercase{\textit{et al.}}: Bare Demo of IEEEtran.cls for Journals}
% The only time the second header will appear is for the odd numbered pages
% after the title page when using the twoside option.
%
% *** Note that you probably will NOT want to include the author's ***
% *** name in the headers of peer review papers.                   ***
% You can use \ifCLASSOPTIONpeerreview for conditional compilation here if
% you desire.

% If you want to put a publisher's ID mark on the page you can do it like
% this:
%\IEEEpubid{0000--0000/00\$00.00~\copyright~2012 IEEE}
% Remember, if you use this you must call \IEEEpubidadjcol in the second
% column for its text to clear the IEEEpubid mark.

% use for special paper notices
%\IEEEspecialpapernotice{(Invited Paper)}

% make the title area
\maketitle

% As a general rule, do not put math, special symbols or citations
% in the abstract or keywords.
\begin{abstract}
The stable spline (SS) kernel and the diagonal correlated (DC) kernel are two kernels that have been applied and studied extensively for kernel-based regularized LTI system identification. In this note, we show that similar to the derivation of the SS kernel, the continuous-time DC kernel can be derived by applying the same ``stable'' coordinate change to a ``generalized'' first-order spline kernel, and thus can be interpreted as a stable generalized first-order spline kernel. This interpretation provides new facets to understand the properties of the DC kernel. In particular, we derive a new orthonormal basis expansion of the DC kernel, and the explicit expression of the norm of the RKHS associated with the DC kernel. Moreover, for the non-uniformly sampled DC kernel, we derive its maximum entropy property and show that its kernel matrix has tridiagonal inverse.

\end{abstract}

% Note that keywords are not normally used for peerreview papers.

% For peer review papers, you can put extra information on the cover
% page as needed:
% \ifCLASSOPTIONpeerreview
% \begin{center} \bfseries EDICS Category: 3-BBND \end{center}
% \fi
%
% For peerreview papers, this IEEEtran command inserts a page break and
% creates the second title. It will be ignored for other modes.
\IEEEpeerreviewmaketitle

% The very first letter is a 2 line initial drop letter followed
% by the rest of the first word in caps.
%
% form to use if the first word consists of a single letter:
% \IEEEPARstart{A}{demo} file is ....
%
% form to use if you need the single drop letter followed by
% normal text (unknown if ever used by IEEE):
% \IEEEPARstart{A}{}demo file is ....
%
% Some journals put the first two words in caps:
% \IEEEPARstart{T}{his demo} file is ....
%
% Here we have the typical use of a "T" for an initial drop letter
% and "HIS" in caps to complete the first word.

\section{INTRODUCTION}

Linear time invariant (LTI) system identification is a classical topic in system identification. The current standard solution to this topic is the maximum likelihood/prediction error method(ML/PEM), see e.g., \cite{Ljung:99}. A new solution is the kernel-based regularization method that is first proposed in \cite{PN10a} and further studied in \cite{PCN11,COL12a,CALCP14}, see \cite{PDCDL14} for a survey of this method.
Recent progress for this method include, e.g., the kernel design \cite{Chen16,ZC17,PCP17},  analysis of the hyper-parameter estimators \cite{PC15,MCL17}, input design \cite{FS17,MC17} and development of the dual theory in frequency domain \cite{LC16}.

This method uses the impulse response model and solves a regularized least squares problem with a suitably designed and tuned kernel. The kernel plays a similar role as the parametric model structure in ML/PEM: on the one hand, it decides in what space the estimated impulse response is searched and on the other hand, it encodes the prior knowledge regarding the underlying system to be identified.
The first two kernels for this method are the stable spline (SS) kernel \cite{PN10a} and the diagonal correlated (DC) kernel \cite{COL12a}. On the one hand, these two kernels are derived in different ways. The SS kernel is obtained by applying a ``stable'' coordinate change to the second-order spline kernel [cf. \eqref{eq:splinekernel-order2}], and the DC kernel is obtained by mimicing the behavior of the optimal kernel \cite[eq. (56)]{COL12a} for this method. On the other hand, these two kernels share some common features \cite{CL14,Chen16}, which are fundamental for developing systematic methods to design kernels for this method. In particular, we have shown in \cite{Chen16} that both kernels belong to the class of amplitude modulated locally stationary kernels, and simulation-induced kernels, leading to a machine learning perspective and a system theory perspective to design more general kernels, respectively.

The SS kernel still has some other features inherited from its mother kernel -- the spline kernel [cf. \eqref{eq:splinekernel}], which were previously regarded as unique for the SS kernel. For example, the orthonormal basis expansion of the SS kernel with respect to a suitably chosen measure can be simply derived by applying the ``stable'' coordinate change to that of the second-order spline kernel \cite{PN10a}, which is key for developing efficient implementation algorithms for this method \cite{CCP:12}. Moreover, the same technique applies when deriving the norm of the reproducing kernel Hilbert space (RKHS) induced by the SS kernel and the maximum entropy property of the SS kernel \cite{CACCLP16}. Interestingly, as will be shown shortly, these features in fact also hold for the continuous-time DC kernel. The key lies in to show that the DC kernel can actually be derived by applying the same ``stable'' coordinate change to a ``generalized'' first-order spline kernel. The DC kernel can thus be interpreted as a stable generalized first-order spline kernel. This interpretation provides new facets to understand the properties of the DC kernel. In particular, we derive a new orthonormal basis expansion of the DC kernel and the explicit expression of the norm of the RKHS associated with the DC kernel. Moreover, for the non-uniformly sampled DC kernel, we derive its maximum entropy property and show that its kernel matrix has tridiagonal inverse, which extend the corresponding results in \cite{CACCLP16,CCL17}.

%%%%%%%%%%%%%%%%%%%%%%%%%%%%%%%%%%%%%%%%%%%%%%%%%%%%%%%%%%%%%%%%%%%%%%%%%%%%%%%%
\section{System identification with Kernel-based Regularization Method}

\subsection{Problem Statement}
We consider continuous-time LTI stable and causal systems, which are described by
\begin{align}
y(t) = (g*u)(t) + v(t),
\end{align} where $t\geq0$ is the time index, $y(t),u(t),v(t)\in\R$ are the measured output, input and
disturbance of the system at time $t$, respectively, $g(t)$ is the impulse
response of the LTI system, and $(g*u)(t)$ is the convolution of
the impulse response $g(t)$ with the input $u(t)$ (evaluated at $t$).
Since $g(t)=0$ for $t<0$ due to the causality assumption, the
convolution $(g*u)(t)$ takes the form of
\begin{align}
&(g*u)(t) =%\nonumber\\& \left\{\begin{aligned}
                  %&
                  \int_{s=0}^{\infty} g(s)u(t-s)ds, %& \text{continuous time (CT)}
%\\
%                  &\sum_{t'=0}^{\infty} g(t')u(t-t') & \text{discrete time (DT)}
%                \end{aligned}
%\right.
\end{align}
%\begin{align}
%g*u(t) = \left\{\begin{array}{cc}
%                  \int_{t'=0}^\infty g(t')u(t-t')dt' & \text{continuous time (CT)}
%\\
%                  \sum_{t'=0}^\infty g(t')u(t-t') & \text{discrete time (DT)}
%                \end{array}
%\right.
%\end{align}
where the unknown input $u(t)$ with $t<0$ is set to zero.  %The disturbance $v(t)$ is assumed to be a
%white Gaussian noise with mean 0 and variance $\sigma^2$ and independent of $u(t)$.
The problem is to estimate $g(t)$ as well as possible based
on the measured data $\{y(t)\}_{t=1}^{N}$ and $u(t)$ with $t\geq0$.

\subsection{Kernel-based Regularization Method}

To estimate the impulse response $g(t)$ from $\{y(t)\}_{t=1}^{N}$ and $u(t)$ with $t\geq0$ is an ill-conditioned problem. To overcome this difficulty, the kernel-based regularization method first introduces a positive semidefinite kernel $k(t,s;\theta)$ with $t,s\geq0$ and constrain the search for a suitable impulse response within the reproducing kernel Hilbert space (RKHS) $\mathcal H_k$ induced by $k(t,s;\theta)$\footnote{The kernel $k(t,s;\theta)$ sometimes will be written as $k(t,s)$ for simplicity. Recall that a function $k:X\times X\rightarrow \R$ with $X$ being a metric space
is called a positive semidefinite kernel, if it is
symmetric and satisfies $\sum_{i,j=1}^m a_ia_j k(x_i,x_j)\geq0$ for
any $m\in\mathbb N$, $\{x_1,\cdots,x_m\}\subset X$ and $\{a_1, ...,
a_m\} \subset \R$. According to the Moore-Aronszajn Theorem, see e.g., \cite{Aronszajn50}, to every positive
semidefinite kernel $k(x,x')$ there exists a unique RKHS $\mathcal H_k$ with $k(x,x')$ as the
reproducing kernel, i.e., with $k_x\triangleq k(x,\cdot)$,
$k(x,x')$ has the
reproducing property $
\inn{f,k_x}_{\mathcal H_k} = f(x)$, for $f\in\mathcal H_k,
x\in X$.},
 where $\theta$ is a hyper-parameter vector that contains the parameters used to parameterize the kernel. In particular, the  regularized least squares criterion is used to estimate $g(t)$:
\begin{align}\label{eq:regulinHilbert} \hat g(t) = \argmin_{g\in\mathcal H_k} \sum_{t=1}^N (y(t) - (g*u)(t))^2 +
\gamma \|g\|_{\mathcal H_k}^2,
\end{align} where $\|\cdot\|_{\mathcal H_k}$ is the norm of $\mathcal H_k$ and  $\gamma>0$ is a regularization parameter and
controls the tradeoff between the data fit  and the regularization term.

The performance of the kernel-based regularization method depends on several factors, such as the choice of the kernel $k(t,s;\theta)$, the estimation of the hyper-parameter $\theta$, and the input design.
Assume that a kernel $k(t,s;\theta)$ has been chosen, an estimate of $\theta$ has been found, and an input $u(t)$ has been designed. Then by setting $\gamma=\sigma^2$ and
%The current most effective method to determine $\theta$ is the so-called empirical Bayes method \cite{PC15}. It first embeds the regutlarization in a Bayesian framework and then estimate $\theta$ and possibly also the noise variance $\sigma^2$ by maximizing the marginal likelihood $p(Y|\eta)$, where $Y\in\R^N$ with $y(t)$ being its $t$th element, and $\eta$ could be $\theta$ or the concatenation of $\theta$ and $\sigma^2$. Specifically,
defining $A\in\R^{N\times N}$ with its $(t,s)$th element $A_{t,s}$ as follows
\begin{align}
a(t,s)=(k(t,\cdot)*u)(s),
A_{t,s}=(a(\cdot,s)*u)(t),
\end{align} %and we then get %\begin{align*}
%  \hat\eta = \argmin_{\eta} Y^T(A+\sigma^2I_N)^{-1}Y + \log\det(A+\sigma^2I_N)
%\end{align*} where $I_N$ is the $N$-dimensional identity matrix. When an estimate of $\eta$ is obtained,
we get the solution
to (\ref{eq:regulinHilbert}) by the representer theorem \cite[Theorem 3, page
671]{PDCDL14}:
%\begin{subequations}
\begin{align}\label{eq:solution2regulinHilbert} \hat g(t)=\sum_{s=1}^N \hat c_{s} a(t,s),
\end{align} where $\hat c_{s}$ is the $s$th element of $\hat c= (A + \gamma I_N)^{-1}Y$. %\begin{align}
%  \hat c = (A + \gamma I_N)^{-1}Y
%\end{align} where $\gamma$ is set to be $\sigma^2$.  \end{subequations}

\subsection{Kernels for Regularized Impulse Response Estimation }

%Apparently, estimating the impulse response $g(t)$ from finite data
%record $\{y(t),u(t)\}_{t=1}^{N}$ is an ill-conditioned problem
%because the solution is not unique. There are at least two ways to
%tackle this problem.
%
%Firstly, note that LTI systems have more compact descriptions in
%terms of transfer functions. Therefore, the problem can be tackled
%by first postulating a finite dimensional parametric transfer
%function and then estimating this transfer function instead. This
%idea leads to the standard method for system identification of LTI
%systems, the so-called maximum likelihood/prediction error method
%(ML/PEM). By use of finite dimensional parametric transfer function,
%the ill-conditioning of the problem is avoided. However, a distinct
%difficulty for ML/PEM is how to choose a suitable model structure
%with the \emph{right} model complexity. This difficulty is often
%dealt with the classical model order selection methods, such as
%Akaike's information criterion and the cross validation methods.

%According to the Moore-Aronszajn Theorem, see e.g., \cite{Aronszajn50}, to every positive
%semidefinite kernel $k(x,x')$ there exists a unique RKHS $\mathcal H_k$ with $k(x,x')$ as the
%reproducing kernel, i.e., with $k_x\triangleq k(x,\cdot)$,
%$k(x,x')$ has the
%reproducing property \begin{align}   \label{eq:rp}
%\inn{f,k_x}_{\mathcal H_k} = f(x), \qquad \forall f\in\mathcal H_k,
%x\in X.
%\end{align} Now we go back to the discussions of the regularized impulse response estimation.

Many kernels
have been introduced, e.g., the stable spline (SS) kernel in
\cite{PN10a} and the tuned correlated (TC) kernel and the diagonal correlated (DC) kernel in
\cite{COL12a}:
\begin{subequations}\label{eq:k-cont}
\begin{align}\label{eq:SS}
&k^{\text{SS}}(t,s;\alpha) = \frac{e^{-\alpha(t+s)}e^{-\alpha\max\{t,s\}}}{2}-\frac{e^{-3\alpha\max\{t,s\}}}{6}, \\
&k^{\text{TC}}(t,s;\beta) = e^{-\beta(t+s)}e^{-\beta|t-s|},\ \beta>0,\label{eq:TC}\\
&k^{\text{DC}}(t,s;\alpha,\beta) = e^{-\alpha(t+s)}e^{-\beta|t-s|},\ \alpha>0,\beta\geq0,\label{eq:DC}
\end{align} where (\ref{eq:TC}) is a special case of (\ref{eq:DC}) with $\alpha=\beta$
\cite{COL12a} and is also called the first-order stable spline (SS-1) kernel. \end{subequations}

The SS kernel (\ref{eq:SS}) and the TC kernel (\ref{eq:TC}) can be derived \cite{PCN11} by applying a ``stable'' coordinate change to the second-order and first-order spline kernel, respectively.
To be specific, we first recall the Sobolev space and its associated kernel \cite{AF03,Mazja13,Wahba:90}. The Sobolev space is the general term used for a functional space whose
norm involves derivatives. The Sobolev space $W_m^0$ defined on
$[0,\ 1]$ is defined as the set of functions $f:[0,\
1]\rightarrow\R$ such that the next conditions hold:
\begin{enumerate}
  \item for $i=0,\cdots,m-1$, its $i$th order derivative $f^{(i)}$ is absolutely continuous\footnote{Let $X$ be an interval of $\R$. A function $f:X\rightarrow \R$ is said to be absolutely continuous on $X$ if for any $\epsilon>0$, there is $\delta >0$ such that when any finite sequence of pairwise disjoint subinterval $(a_i,b_i)$ of $X$ satisfies $\sum_i (a_i-b_i)<\delta$, then $\sum_i |f(a_i)-f(b_i)|<\epsilon$. Moreover, $f$ is absolutely continuous on a compact interval $[a, b]$ if and only if $f$ has a derivative $f^{(1)}$ almost everywhere and $f^{(1)}$ is Lebesgue integrable, and $f(x)=f(a)+\int_a^x f^{(1)}(\tau)d\tau$.}  and moreover, $f^{(i)}(0)=0$;
  \item its $m$th order derivative $f^{(m)}\in L^2([0,1])$.
\end{enumerate}
The inner product on $W_m^0$ over $\R$ is defined through the
classical inner product in $L^2([0,1])$:
\begin{align}
\langle f,h \rangle_{W_m^0} &= \langle f^{(m)},h^{(m)}
\rangle_{L^2([0,1])}\nonumber\\ &=\int_0^1 f^{(m)}(\tau)h^{(m)}(\tau) d\tau.
\end{align}
It can be verified  \cite{AF03,Mazja13,Wahba:90} that $W_m^0$ is a RKHS
with the reproducing kernel, often called the spline kernel,
\begin{align}\label{eq:splinekernel} &w_m^{\text{S}}(\tau,\nu) =
\int_0^1 G_m(\tau,s) G_m(\nu,s)ds,\  \tau,\nu\in[0,1],
\end{align} %and the norm \begin{align}
%\|f\|^2_{W_m^0}=\int_0^1 (f^{(m)}(t))^2dt
%\end{align}
where the Green's function
$G_m(\tau,s)=(\tau-s)_+^{m-1}/(m-1)!$ with $(x)_+=x$ for $x\geq0$
and $(x)_-=0$ for $x<0$. In particular, the spline kernels (\ref{eq:splinekernel}) with $m=1,2$ are
called the first-order and second-order spline kernel, respectively:
\begin{align}\label{eq:splinekernel-order1}
&w_1^{\text{S}}(\tau,\nu) = \min\{\tau,\nu\},\ \tau,\nu\in[0,1],\\
&w_2^{\text{S}}(\tau,\nu) = \frac{1}{2}\tau\nu\min\{\tau,\nu\}-\frac{1}{6}(\min\{\tau,\nu\})^3,\ \tau,\nu\in[0,1].\label{eq:splinekernel-order2}\end{align}
Then it is easy to verify the following result.

\begin{Proposition}[{\cite{PCN11}}]\label{prop:sstc_stablespline}
Consider the SS kernel (\ref{eq:SS}) and the TC kernel (\ref{eq:TC}). Then the following result holds
\begin{align}
  &k^{\text{SS}}(t,s;\alpha) =  w_2^{\text{S}}(e^{-\alpha t},e^{-\alpha s}),\\
  &k^{\text{TC}}(t,s;\beta) =  w_1^{\text{S}}(e^{-2\beta t},e^{-2\beta s}).
  \end{align}
\end{Proposition}

It follows from Proposition \ref{prop:sstc_stablespline} that the SS and TC kernels have several properties inherited from their mother kernel (\ref{eq:splinekernel}), such as the orthonormal basis expansion, the explicit expression of the norm, and the maximum entropy property. Interestingly,
the DC kernel can be derived by applying the same ``stable'' coordinate change to a ``generalized'' first-order spline kernel so that
the DC kernel have similar properties as the SS and TC kernels inherited from the spline kernel (\ref{eq:splinekernel-order1}).

\section{Stable Generalized first-order Spline Kernel}

To show this, we recall the generalized Sobolev space and its associated kernel  \cite{AF03,Mazja13,Wahba:90}.
The generalized Sobolev space is derived by replacing the $i$th order derivatives $i=1,\cdots,m$, in the definition of $W_m^0$,
by more general derivatives. Specifically, the generalized $m$th order derivative of $f$ are defined as follows:
\begin{align} D_m f = \frac{d }{d\tau}\frac{1}{a_1} \frac{d }{d\tau}\frac{1}{a_2}\cdots \frac{d }{d\tau}\frac{1}{a_m}  f,
\end{align}
where the functions $a_i(\tau)$ with $\tau\in[0,\ 1]$, $i=1,\cdots,m$ are functions such that
$D_m f$ is well-defined, and moreover,  \begin{align}\nonumber M_0 f
&= f, M_1 f = \frac{d }{d\tau}\frac{1}{a_m} f, M_2 f = \frac{d
}{d\tau}\frac{1}{a_{m-1}}\frac{d }{d\tau}\frac{1}{a_m} f,\\&\cdots,
M_{m-1}f = \frac{d }{d\tau}\frac{1}{a_2}\cdots \frac{d
}{d\tau}\frac{1}{a_m} f.
\end{align}
Then the generalized Sobolev space $\tilde W_m^0$ defined on $[0,\
1]$ is defined as the set of functions $f:[0,\ 1]\rightarrow\R$ such
that the following conditions hold:
\begin{enumerate}
  \item for $i=0,\cdots,m-1$, $M_i f$ is absolutely continuous and moreover, $M_if(0)=0$;
  \item $D_mf\in L^2([0,1])$.
\end{enumerate}

Note that the functions $f:[0,\ 1]\rightarrow\R$ such that
$M_if(0)=0$, $i=0,\cdots,m-1$ can be represented in the following
form
\begin{subequations}\label{eq:f}
\begin{align}
f(\tau) &= a_m(\tau) \int_0^\tau (D_m f)(s)ds, \ m=1,\\
f(\tau) &= a_m(\tau)
\int_0^\tau a_{m-1}(t_{m-1})dt_{m-1}\int_0^{t_{m-1}}\nonumber\\&
a_{m-2}(t_{m-2}) dt_{m-2}\cdots \int_0^{t_1}(D_m f)(s)ds,\ m>1.
\end{align}
\end{subequations}
By interchanging the integration order in (\ref{eq:f}), we have
\begin{align}
f(\tau) &= \int_0^\tau \tilde G_m(\tau,s)(D_m f)(s)ds
\end{align} where
\begin{subequations}
\begin{align}
\tilde G_m(\tau,s)&=\left\{\begin{array}{cc}
                                   0 & \tau\leq s\\
                                   a_m(\tau) & \tau>s
                                 \end{array}
\right.,\ m=1,\\
\tilde G_m(\tau,s) &= a_m(\tau) \int_s^\tau a_1(t_1)dt_1\int_{t_1}^\tau \nonumber\\
&a_2(t_2) dt_2\cdots \int_{t_{m-2}}^\tau a_{m-1}(t_{m-1})dt_{m-1},\ m>1.
\end{align}
\end{subequations}
Moreover, the inner product on $\tilde W_m^0$ over $\R$ is defined
through the classical inner product in $L^2([0,1])$:
\begin{align}\label{eq:tildewm0-innerprod}
\langle f,h \rangle_{\tilde W_m^0} &= \langle D_m f, D_m h
\rangle_{L^2([0,1])}\nonumber\\& =\int_0^1 (D_mf)(\tau)(D_mh)(\tau) d\tau.
\end{align}
It can be verified  \cite{AF03,Mazja13,Wahba:90} that the generalized Sobolev space $\tilde W_m^0$
is a RKHS with the reproducing kernel
\begin{align}\label{eq:gsplinekernel} & w_{m}^{\text{GS}}(\tau,\nu)
= \int_0^1 \tilde G_m(\tau,s)\tilde G_m(\nu,s)ds,
\end{align} which is called the generalized spline kernel here.
% and the norm \begin{align}
%\|f\|^2_{\tilde W_m^0}=\int_0^1 (D_m f)^2(t)dt
%\end{align}

Then the following result can be proved for the DC kernel\footnote{All proofs of the propositions are deferred to the Appendix.}.

\begin{Proposition}\label{prop:dc_spline} Consider the DC kernel (\ref{eq:TC})  and the generalized spline kernel (\ref{eq:gsplinekernel}).
If we take \begin{align}\label{eq:assumption} m=1,\
a_1(\tau)=\tau^\rho,\ \rho> -0.5,
\end{align} then we have
\begin{align}
k^{\text{DC}}&(t,s;\alpha,\beta)=w_1^{\text{GS}}(e^{-2\beta
t},e^{-2\beta
s};\frac{\alpha-\beta}{2\beta}),\label{eq:dc_gs2}\end{align}
where
\begin{equation}\begin{aligned}\label{eq:gsplinekernel-order1}
&w_1^{\text{GS}}(\tau,\nu;\rho) = \tau^\rho \nu^\rho\min\{\tau,\nu\},\ \tau,\nu\in[0,1].
\end{aligned}\end{equation}
\end{Proposition}

%The kernel (\ref{eq:gsplinekernel}) is closely related with
%the spline kernel (\ref{eq:splinekernel}) and can thus be called a
%generalized spline kernel.
%In particular, if we take \begin{align}\label{eq:assumption} m=1,\
%a_m(\tau)=\tau^\rho,\ \rho> -0.5
%\end{align} then we get the generalized first-order spline kernel
%\begin{equation}\begin{aligned}\label{eq:gsplinekernel-order1}
%&w_1^{\text{GS}}(\tau,\nu;\rho) = \tau^\rho \nu^\rho\min\{\tau,\nu\},\ \tau,\nu\in[0,1].
%\end{aligned}\end{equation}
%Its implication is shown in Fig.
%\ref{fig:dc_gspline}, where $h(\tau)$ is a zero mean Gaussian process
%with the first-order spline kernel (\ref{eq:splinekernel-order1}) being its covariance
%function and the block $\tau^\rho$ represents a pointwise scale factor
%of  $h(\tau)$, and it is straightforward to verify that
%the output Gaussian process has zero mean and the generalized first-order spline
%kernel (\ref{eq:gsplinekernel-order1}) as its covariance function.
%\begin{figure}
%\begin{center}
%\includegraphics[width=2.5in, height=1in]{cdc16_2}
%\end{center}
%\caption{A graphical interpretation of the generalized first-order spline kernel (\ref{eq:gsplinekernel-order1}) and its connection with the first-order spline kernel (\ref{eq:splinekernel-order1})}\label{fig:dc_gspline}
%\vspace{-4mm}\end{figure}

The kernel (\ref{eq:gsplinekernel-order1}) is called the generalized first-order spline kernel here.
Recall from Proposition \ref{prop:sstc_stablespline} that the SS kernel (\ref{eq:SS}) is called the
stable spline kernel, because it can be obtained by
applying a ``stable'' coordinate change to the second-order spline kernel. Now Proposition \ref{prop:dc_spline} shows that
the DC kernel (\ref{eq:DC})
can be obtained by applying the same ``stable'' coordinate change
to the generalized first-order spline kernel (\ref{eq:gsplinekernel-order1}).
The DC kernel can thus be called a stable
generalized first-order spline kernel. As will be shown below, this finding provides new facets to understand the properties of
the DC kernel.

%\begin{Remark}\label{rmk:rhovalue}
%Recall from e.g., \cite[Theorem 3.1]{Astrom:70} that a sufficient condition for a positive semidefinite kernel $k(x,x')$ to be continuous is that $k(x,x)$ is continuous. Therefore, we impose $\rho> -0.5$ in (\ref{eq:assumption}) to guarantee  (\ref{eq:gsplinekernel-order1}) is continuous. Note that $\rho = \frac{\alpha-\beta}{2\beta}$ with $\alpha>0$ and $\beta\geq0$, thus $\rho>-0.5$, which satisfies the assumption made in (\ref{eq:assumption}).
%\end{Remark}

\section{Norm and Orthonormal Basis Expansion}

In this section, we derive the norm and the
orthonormal basis expansion of the generalized first-order spline kernel and the DC kernel from their mother kernel (\ref{eq:splinekernel-order1}).

To this goal, we first recall that when $X$ is compact and the positive
semidefinite kernel $k(x,x')$ with $x,x'\in X$ is continuous, $\mathcal H_k$ has the following  property by Mercer's Theorem, see e.g., \cite{Mercer09}, \cite[Thm. 17, page 90]{Hochstadt73}, \cite[Thm. 1, page 34]{CS02}.

Let $\mu$ be a Borel measure on $X$ and $L^2(X,\mu)$\footnote{When $X$ is a subset of $\R$ and $\mu$ is the Lebesgue measure, $L^2(X,\mu)$ will be written as $L^2(X)$ for simplicity.} be the space of functions $f$ for which $\int_X (f(x))^2d\mu(x)<\infty$. Then the integral operator $L_k\phi(x)$ on $L^2(X,\mu)$:
\begin{align}\label{eq:integral_op} L_k\phi(x)\triangleq
\int_X k(x,x')\phi(x')d\mu(x'), \ x\in X.
\end{align} has at most countably many positive eigenvalues $\{\lambda_i\}_{i=1}^\infty$ and orthonormal eigenfunctions $\{\phi_i\}_{i=1}^\infty$
\footnote{If for some $\lambda$, the homogenous integral equation
\begin{align}\label{eq:integral_def} L_k\phi(x)=\lambda \phi(x), \ x\in X
\end{align} has solutions other than $\phi(x)=0$,  $\lambda$ and the solutions of (\ref{eq:integral_def}) are
called the eigenvalues and eigenfunctions of the integral operator
$L_k$, respectively.
}, and the positive semidefinite kernel $k(x,x')$ has a series expansion
\begin{align}
k(x,x')=\sum_{i=1}^\infty \lambda_i \phi_i(x)\phi_i(x'),
\end{align}
which converges uniformly and absolutely on $X\times X$. Moreover,  $\{\sqrt{\lambda_i}\phi_i\}_{i=0}^\infty$ forms an orthonormal basis of $\mathcal H_k$, which gives an alternative representation of $\mathcal H_k$ by \cite[Thm. 4]{CS02}: \begin{align}
\mathcal H_k = \{f:f=\sum_{i=1}^\infty f_i \phi_i, \text{with } \sum_{i=1}^\infty\frac{f_i^2}{\lambda_i}<\infty\}.
\end{align}

Now we recall the eigenvalues and the orthonormal eigenfunctions of the first-order spline kernel (\ref{eq:splinekernel-order1}) on $L_2([0,\ 1])$, which are well-known, see e.g., \cite[equations
(119)-(120), page 196]{Trees01}:
\begin{align}\label{eq:splinekernel-order1-expansion}
\int_0^1 & \min\{\tau,\nu\}\varphi_i(\nu)d\nu = \lambda_i\varphi(\tau),\ i=1,2,\cdots \nonumber\\
\text{where }\lambda_i&=\frac{1}{(i-\frac{1}{2})^2\pi^2},\varphi_i(\tau)
= 2^{\frac{1}{2}}\sin((i-\frac{1}{2})\pi \tau).\end{align}
Moreover, Mercer's Theorem \cite{Mercer09}, \cite[Thm. 17, page 90]{Hochstadt73}, \cite[Thm. 1, page 34]{CS02}
guarantees  that the series
\begin{align}\label{eq:splinekernel-order1-expansion2}
\min&\{\tau,\nu\}=\sum_{i=1}^\infty\lambda_i\varphi_i(\tau)\varphi_i(\nu)
\end{align}
converges absolutely and uniformly on $[0,\ 1]\times [0,\ 1]$, and $\{\sqrt{\lambda_i}\varphi_i\}_{i=1}^\infty$ forms an orthonormal basis of $W_1^0$ and  $W_1^0$ has an equivalent representation: \begin{align}
W_1^0=\{h|h(\tau)=\sum_{i=1}^\infty h_i\varphi_i(\tau), \tau\in[0,\ 1],
\sum_{i=1}^\infty \frac{h_i^2}{\lambda_i}<\infty\},
\end{align} and the norm of $h$ can be computed according to
\begin{align}
\|h\|_{W_1^0}^2 = \sum_{i=1}^\infty \frac{h_i^2}{\lambda_i}\left(=\int_0^1 (h^{(1)}(\tau))^2 d\tau\right)
\end{align}

Noticing the similarity between (\ref{eq:splinekernel-order1}) and (\ref{eq:gsplinekernel-order1}), we can prove the next result for the generalized first-order spline kernel (\ref{eq:gsplinekernel-order1}).
\begin{Proposition}
\label{prop:gs-obs} Consider the generalized first-order spline kernel (\ref{eq:gsplinekernel-order1}). Let the eigenvalues and orthonormal eigenfunctions of the first-order spline kernel (\ref{eq:splinekernel-order1}) take the form of (\ref{eq:splinekernel-order1-expansion}). Then the following results hold:
\begin{enumerate}
\item Let $\phi_i(\tau)=\tau^\rho\varphi_i(\tau)$, $i=1,2,\cdots$. Then $\{\lambda_i\}_{i=1}^\infty$ and $\{\phi_i\}_{i=1}^\infty$ are
the eigenvalues and orthonormal eigenfunctions of the generalized first-order spline kernel (\ref{eq:gsplinekernel-order1}) on $L_2([0,\ 1],\mu(\nu))$ with the measure $\mu(\nu)$ such that $d\mu(\nu)=\nu^{-2\rho}d\nu$, respectively.

\item The series
\begin{align}\label{eq:gsplinekernel-order1-expansion}
\tau^\rho\nu^\rho\min&\{\tau,\nu\}=\sum_{i=1}^\infty\lambda_i\phi_i(\tau)\phi_i(\nu)
\end{align}
converges absolutely and uniformly on $[0,\ 1]\times [0,\ 1]$.

\item $\{\sqrt{\lambda_i}\phi_i\}_{i=1}^\infty$ forms an orthonormal basis of $\tilde W_1^0$, and
$\tilde W_1^0$ has an equivalent representation:
\begin{align}\label{eq:rkhs_gs_altn}
\tilde W_1^0=\{f|f(\tau)=&\sum_{i=1}^\infty f_i\phi_i(\tau),\nonumber\\ &\tau\in[0,1],
\sum_{i=1}^\infty \frac{f_i^2}{\lambda_i}<\infty\},
\end{align} and the norm of $f$ can be computed according to
\begin{align}
\|f\|_{\tilde W_1^0}^2 = \sum_{i=1}^\infty \frac{f_i^2}{\lambda_i}\left(=\int_0^1 \left(\frac{d}{d\tau} \frac{f(\tau)}{\tau^\rho}\right)^2 d\tau\right).
\end{align}
 %Let $f_i=\langle f,\phi_i\rangle_{\tilde W_1^0}$. Then

\end{enumerate}
\end{Proposition}

By Proposition \ref{prop:gs-obs} and noting (\ref{eq:dc_gs2}), we can prove the next result for the DC kernel (\ref{eq:DC}).
\begin{Proposition}
\label{prop:dc-obs} Consider the DC kernel (\ref{eq:DC}). Let the eigenvalues and orthonormal eigenfunctions of the generalized first-order spline kernel (\ref{eq:gsplinekernel-order1}) take the form of (\ref{eq:gsplinekernel-order1-expansion}). Then the following results hold:
\begin{enumerate}
\item Let $\psi_i(t)=\phi_i(e^{-2\beta t})$, $i=1,2,\cdots$. Then $\{\lambda_i\}_{i=1}^\infty$ and $\{\psi_i\}_{i=1}^\infty$
are the eigenvalues and the orthonormal eigenfunctions of the DC kernel (\ref{eq:DC}) on $L_2([0,\ \infty),\iota(t))$ with the measure $\iota(t)$ such that $d\iota(t)=2\beta e^{2\beta(2\rho-1) t}dt$, respectively.

\item The series
\begin{align}\label{eq:DC_obs_NotLebesgue}
e^{-\alpha(t+s)}e^{-\beta|t-s|}=\sum_{i=1}^\infty\lambda_i\psi_i(t)\psi_i(s)
\end{align}
converges absolutely and uniformly on $Z\times Z'$ with $Z,Z'$ being any compact subsets of $[0,\ \infty)$.

\item $\{\sqrt{\lambda_i}\psi_i\}_{i=1}^\infty$ forms an orthonormal basis of the RKHS $\mathcal H^{\text{DC}}$ induced by the DC kernel (\ref{eq:DC}), and
$\mathcal H^{\text{DC}}$ has an equivalent representation: \begin{align}\label{eq:rkhs_dc_altn}
\mathcal H^{\text{DC}}=\{g|g(t)=\sum_{i=1}^\infty g_i\psi_i(t), t\geq0,
\sum_{i=1}^\infty \frac{g_i^2}{\lambda_i}<\infty\},
\end{align} and the norm of $g$ can be computed according to
\begin{align}
\|g\|_{\mathcal H^{\text{DC}}}^2 = \sum_{i=1}^\infty \frac{g_i^2}{\lambda_i}.
\end{align}

\item $\mathcal H^{\text{DC}}$ and $\tilde W_1^0$ are isometrically isomorphic and
\begin{align}\label{eq:dc_norm}
\|g\|_{\mathcal H^{\text{DC}}}^2 = \int_0^\infty 2\beta e^{(4\rho+2)\beta t}\left(\frac{1}{2\beta} g^{(1)}(t) + \rho g(t) \right)^2dt.
\end{align}
\end{enumerate}
\end{Proposition}

\begin{Remark} It should be noted that the orthonormal basis expansion of the DC kernel with respect to the \emph{Lebesgue measure} has been derived before in \cite{CPCL15cdc} and is different from (\ref{eq:DC_obs_NotLebesgue}).
Then a natural question is "what is their difference?". It turns out that these orthonormal basis expansions are optimal in some sense with respect to certain kinds of inputs. Due to the limitation of the space, the details cannot be put here but will be in an independent paper.
\end{Remark}

\begin{Remark} When $\rho=0$, the DC kernel (\ref{eq:DC}) reduces to the TC kernel (\ref{eq:TC}).
In this case, we have \begin{align}\label{eq:tc_norm}
\|g\|_{\mathcal H^{\text{TC}}}^2 = \int_0^\infty \frac{1}{2\beta} e^{2\beta t}\left(g^{(1)}(t)\right)^2dt
\end{align} Comparing (\ref{eq:tc_norm}) with (\ref{eq:dc_norm}) shows from another perspective that the DC kernel is more flexible than the TC kernel for regularize impulse response estimation. For illustration, if $g(t)$ is constrained to be $g(t)=e^{-\gamma t}$, then a necessary condition for $g\in \mathcal H^{\text{TC}}$ is  $\gamma>\beta$, but a necessary condition for $g\in \mathcal H^{\text{DC}}$ is  $\gamma >(2\rho+1)\beta$, where $\rho> -0.5$.

\end{Remark}

%On the one hand, note from (\ref{eq:tildewm0-innerprod}) that
%\begin{align}\label{eq:tildewm0-norm} \|f\|_{\tilde W_1^0}^2 =
%\int_0^1 \frac{d}{dt}\left(\frac{f(t)}{t^\rho}\right) dt
%\end{align} On the other hand,

\section{Maximum Entropy Interpretation of Non-uniformly Sampled DC Kernel}\label{sec:maxent_dc_sgs}

In this section, for the non-uniformly sampled DC kernel (\ref{eq:DC}), we first derive its maximum entropy (MaxEnt) property from its mother kernel (\ref{eq:splinekernel-order1}) using arguments similar to \cite{CACCLP16} and then show that its kernel matrix has tridiagonal inverse. %It should be noted that the latter result is more general than the one in \cite{CCL17}, because the continuous-time DC kernel was uniformly sampled there but non-uniformly here.
%In particular, we will sample the kernel and consider stochastic processes defined on an ordered index set
%$\mathcal{T}=\{t_i| 0\leq t_i<t_{i+1}\leq\infty,i=0,1,2,\cdots,\}$.

First, recall that a real-valued stochastic process $w(i)$ with $i=0,1,2,\cdots,$ is called a white
Gaussian noise if the $w(i)$'s are independent identically Gaussian
distributed with zero mean and constant variance\footnote{Without loss of generality, we assume the variance is 1.}. Then we construct a Gaussian process
$f(\tau)$ defined on an ordered index set
$\Gamma=\{\tau_i| 0\leq \tau_i<\tau_{i+1}\leq1,i=0,1,2,\cdots,\}$ as follows:
\begin{equation}\label{eq:gp-gsplinekernel-order}\begin{aligned} f(\tau_0)&=0\text{ with
}\tau_0=0,\\ f(\tau_k) &= \tau_k^\rho\sum_{i=1}^k w(i-1) \sqrt{\tau_i-\tau_{i-1}},
k=1,2,\cdots.\end{aligned}\end{equation}
It is easy to verify that $f(\tau)$ with $\tau\in\Gamma$ has the generalized first-order spline kernel (\ref{eq:gsplinekernel-order1}) with $\tau,\nu\in\Gamma$ as its covariance function.
Then we can prove the next result for the kernel (\ref{eq:gsplinekernel-order1}).
\begin{Proposition}
\label{prop:maxent_gsplinekernel-order1}
Let $h(\tau)$ with $\tau\in\Gamma$ be any stochastic process with
$h(\tau_0)=0$ for $\tau_0=0$. For any $n\in\mathbb N$, the stochastic
process $f(\tau)$ in (\ref{eq:gp-gsplinekernel-order}) is the optimal solution to the MaxEnt\footnote{Recall that for a real-valued random variable $X$, the differential entropy $H(X)$ of $X$ is defined as $
H(X)=-\int_S{p(x)\log{p(x)} \text{d} x}$, where $p(x)$ is the probability density function of $X$ and $S$ is the support set
of $X$.} problem
\begin{align}\nonumber
\maximize_{h(\cdot)} &\quad H(h(\tau_1),h(\tau_2),\cdots,h(\tau_{n}))\\
&\text{\emph{subject to}}\nonumber\quad    \mean{h(\tau_i)}=0,\ i=1,\cdots, n\\
&\var{\frac{h(\tau_{1})}{\tau_1^\rho}}=
\tau_1,\label{eq:maxent4gsplinekernel-order1}\\\nonumber& \var{\frac{h(\tau_{i})}{\tau_i^\rho}-\frac{h(\tau_{i-1})}{\tau_{i-1}^\rho}}=
\tau_i-\tau_{i-1},i=2,\cdots,n
\end{align}
where $\mean{\cdot}$ and $\var{\cdot}$ represent the expectation and variance, respectively, and for simplicity, $H(h(t_1),h(t_2),\cdots,h(t_{n}))$ denotes the differential entropy of
$[h(t_1)\ h(t_2)\ \cdots\ h(t_{n})]^T$.
\end{Proposition}

Based on the stochastic
process $f(\tau)$ in (\ref{eq:gp-gsplinekernel-order}), we define another Gaussian process $g(t)$ defined on an ordered index set
$\mathcal{T}=\{t_i| 0\leq t_i<t_{i+1}\leq\infty,i=0,1,2,\cdots,\}$ as follows:
\begin{align}\nonumber g(t_k) &= e^{-2\beta\rho t_k}\sum_{i=k}^{n-1} w(n-1-i) \sqrt{e^{-2\beta t_i} - e^{-2\beta t_{i+1}}},\\& k=0,\cdots,n-1,
g(t_n) =0 \text{ with } t_n=\infty.\label{eq:sol2maxent4dc}
\end{align}
It is easy to verify that $g(t)$ with $t\in\mathcal T$ has the DC kernel (\ref{eq:DC}) with $t,s\in\mathcal{T}$ as its covariance function.
Then we can prove the next result for the DC kernel (\ref{eq:DC}).
\begin{Proposition}
\label{prop:maxent_dc}
Let $h(t)$  be any stochastic process with $h(t_n)=0$ for
$t_n=\infty$. For any $n\in\mathbb N$, the stochastic
process $g(t)$ in (\ref{eq:sol2maxent4dc}) is the optimal solution to the MaxEnt problem
\begin{align}
&\maximize_{h(\cdot)} \quad H(h(t_0),h(t_1),\cdots,h(t_{n-1}))\nonumber\\
&\mbox{subject to}\quad
\mean{h(t_i)}=0, i=0,\cdots, {n-1}\nonumber\\\label{eq:maxentdc}&
\var{\frac{h(t_{i+1})}{e^{-2\beta\rho t_{i+1}}}-\frac{h(t_{i})}{e^{-2\beta\rho t_{i}}}}=
e^{-2\beta t_i}-e^{-2\beta t_{i+1}},\\&\qquad\qquad\nonumber i=0,1,\cdots,n-2\\\nonumber&
\var{\frac{h(t_{n-1})}{e^{-2\beta\rho t_{n-1}}}}= e^{-2\beta t_{n-1}}.
\end{align}
\end{Proposition}

Proposition \ref{prop:maxent_dc} leads to an interesting result that, the kernel matrix of the DC kernel (\ref{eq:DC}) with $t,s\in\mathcal{T}$ has tridiagonal inverse, which is an extension of the result of \cite{CACCLP16} from the TC kernel (\ref{eq:TC}) to the DC kernel (\ref{eq:DC}) and an extension of the result of \cite{CCL17} from the uniformly sampled case to the non-uniformly sampled case.

\begin{Proposition}\label{prop:bidiaginv}
Consider the DC kernel (\ref{eq:DC}) with $t,s\in\mathcal{T}$. Then its kernel matrix $K^{\text{DC}}\in\R^{\bar n\times \bar n}$ with $\bar n\geq 3$ has tridiagonal inverse, where the $(i,j)$th element of $K^{\text{DC}}$ is equal to $k^{\text{DC}}(t_i,t_j;\alpha,\beta)$ with $t_i,t_j\in\mathcal T$.
  \end{Proposition}

\begin{figure}[!t]
\begin{center}
\includegraphics[width=2in, height=1.5in]{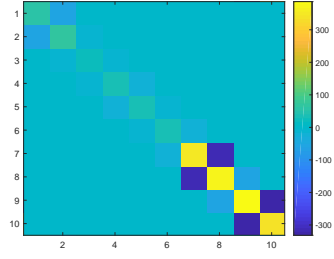}
\end{center}
\caption{Scaled image of $(K^{\text{DC}})^{-1}$ in Example \ref{exm:DC_bandedinverse}. The image is drawn by using \texttt{imagesc} in MATLAB, where the colder the color the smaller the element of $(K^{\text{DC}})^{-1}$.}
\label{fig:dc_bandedinverse}
\end{figure}

For illustration, we consider an example.
\begin{Example}\label{exm:DC_bandedinverse}
We consider the inverse of $K^{\text{DC}}\in\R^{\bar n\times \bar n}$ with $\bar n=10$, $\alpha=-0.2$, $\beta=-0.3$ and $t_i,t_j$ with $i,j=1,\dots,10$ take values from the set generated in Matlab with  \texttt{sort(rand(10,1))}. By Proposition \ref{prop:bidiaginv}, the inverse of $K^{\text{DC}}$ should be tridiagonal, which is confirmed by
Fig. \ref{fig:dc_bandedinverse}.
\end{Example}

\section{Conclusion}

In this note, we have shown that the continuous-time diagonal correlated (DC) kernel can be interpreted as a stable generalized first-order spline kernel. This interpretation provides new facets to understand the properties of the DC kernel. In particular, we derive a new orthonormal basis expansion of the DC kernel and the explicit expression of the norm of the RKHS associated with the DC kernel. Moreover, for the non-uniformly sampled DC kernel, we derive its maximum entropy property and show that its kernel matrix has tridiagonal inverse. There are several interesting works that can be done in the future. For example, it is interesting and possible to derive the explicit expression of the norm of the sampled DC kernel for discrete time linear time invariant system identification, and compare the derived orthonormal basis expansion with the one in \cite{CPCL15cdc}.

\appendix

\subsection{Proof of Proposition \ref{prop:dc_spline}}

On the one hand, it is easy to check that under the assumption (\ref{eq:assumption}) the generalized spline kernel (\ref{eq:gsplinekernel}) takes the form of (\ref{eq:gsplinekernel-order1}). On the other hand, the DC kernel (\ref{eq:DC}) can be rewritten as follows:
\begin{align*} k^{\text{DC}}&(t,s;\alpha,\beta)=e^{(\beta-\alpha)(t+s)} k^{\text{TC}}(t,s;\beta)
 \\&= \left(e^{-2\beta t}\right)^{-\frac{\beta-\alpha}{2\beta}}\left(e^{-2\beta s}\right)^{-\frac{\beta-\alpha}{2\beta}}\min\{e^{-2\beta t},e^{-2\beta s}\}.
\end{align*} Then noting the above equation and (\ref{eq:gsplinekernel-order1}) gives the result.

\subsection{Proof of Proposition \ref{prop:gs-obs}}

We first consider the proof for 1). For any $i\in\mathbb N$, we have
\begin{align*}
\int_0^1&  \tau^\rho\nu^\rho\min\{\tau,\nu\} \phi_i(\nu)d\mu(\nu)\\& =
\int_0^1 \tau^\rho\nu^\rho\min\{\tau,\nu\} \nu^\rho\varphi_i(\nu)\nu^{-2\rho}d\nu\\ &=
\lambda_i\tau^\rho \varphi_i(\tau) = \lambda_i\phi_i(\tau)
\end{align*} and moreover, note that the orthonormality of $\{\phi_i\}_{i=1}^\infty$ in
$L_2([0,\ 1],\mu(\nu))$ follows from that of $\{\varphi_i\}_{i=1}^\infty$ in
$L_2([0,\ 1])$.

For 2) and 3), we first show that the kernel (\ref{eq:gsplinekernel-order1}) is continuous. By Schwartz inequality, we have \begin{align*}
&|w_1^{\text{GS}}(\tau+h,\nu+k) - w_1^{\text{GS}}(\tau,\nu) |
 \\&
\leq \{w_1^{\text{GS}}(\tau+h,\tau+h)\times\\&\qquad [w_1^{\text{GS}}(\nu+k,\nu+k)-2w_1^{\text{GS}}(\nu+k,\nu) + w_1^{\text{GS}}(\nu,\nu) ]\}^{\frac{1}{2}}\\
&\quad+ \{w_1^{\text{GS}}(\nu,\nu)\times\\&\qquad [w_1^{\text{GS}}(\tau+h,\tau+h)-2w_1^{\text{GS}}(\tau+h,\tau) + w_1^{\text{GS}}(\tau,\tau) ]\}^{\frac{1}{2}}.
\end{align*} Now we let $h,k\rightarrow 0$, then $w_1^{\text{GS}}(\nu+k,\nu+k)\rightarrow w_1^{\text{GS}}(\nu,\nu)$ because $w_1^{\text{GS}}(\tau,\nu)$ is continuous along the diagonal $\tau=\nu$ and
$w_1^{\text{GS}}(\nu+k,\nu)\rightarrow w_1^{\text{GS}}(\nu,\nu)$ because $w_1^{\text{GS}}(\tau,\nu)$ is continuous in $\tau$. This means that the right-hand side of the above inequality converges to zero as $h,k\rightarrow 0$ and thus the kernel (\ref{eq:gsplinekernel-order1}) is continuous. Then noting that the measure $\mu(\nu)$ is a Borel measure, the results follow from \cite[Thm. 1, page 34]{CS02} and \cite[Thm. 4, page 37]{CS02}, respectively.
This completes the proof.

\subsection{Proof of Proposition \ref{prop:dc-obs}}

The proof 1) to 3) can be done by applying the coordinate change $\tau = e^{-2\beta t}$ to 1) to 3) of Proposition \ref{prop:gs-obs}.
So we only prove 4) below.

Given any $f\in\tilde W_1^0$, it follows from (\ref{eq:rkhs_gs_altn})
that there exist $f_i$, $i=1,2,\cdots,$ such that $f(\tau)=\sum_{i=1}^\infty f_i \phi_i(\tau)$ with $\tau\in[0,\ 1]$.
Taking $\tau = e^{-2\beta t}$, we get a unique $g(t) \triangleq f(e^{-2\beta t})=\sum_{i=1}^\infty f_i \phi_i(e^{-2\beta t})=\sum_{i=1}^\infty f_i \psi_i(t)\in \mathcal H^{\text{DC}}$ since $\sum_{i=1}^\infty f_i^2/\lambda_i<\infty$, and moreover, $\|f\|_{\tilde W_1^0}^2=\|g\|_{\mathcal H^{\text{DC}}}^2=\sum_{i=1}^\infty f_i^2/\lambda_i$. Similarly, given any $g\in\mathcal H^{\text{DC}}$, it follows from (\ref{eq:rkhs_dc_altn})
that there exist $g_i$, $i=1,2,\cdots$ such that $g(t)=\sum_{i=1}^\infty g_i \psi_i(t)$ with $t\geq0$. Taking $t=\log\tau/(-2\beta)$, we get a unique $f(\tau) \triangleq g(\log\tau/(-2\beta))=\sum_{i=1}^\infty g_i \psi_i(\log\tau/(-2\beta))=\sum_{i=1}^\infty g_i \phi_i(\tau)\in \tilde W_1^0$,  since $\sum_{i=1}^\infty g_i^2/\lambda_i<\infty$, and moreover, $\|g\|_{\mathcal H^{\text{DC}}}^2=\|f\|_{\tilde W_1^0}^2=\sum_{i=1}^\infty g_i^2/\lambda_i$. Therefore, $\mathcal H^{\text{DC}}$ and $\tilde W_1^0$ are isometrically isomorphic. Given any $g\in\mathcal H^{\text{DC}}$, we have
\begin{align*}
\|g\|_{\mathcal H^{\text{DC}}}^2&=\|f\|_{\tilde W_1^0}^2 =\int_0^1 \left(\frac{d}{d\tau} \frac{f(\tau)}{\tau^\rho}\right)^2 d\tau\\
&=\int_0^1 \left(\frac{d}{d\tau} \frac{g(\frac{\log\tau}{-2\beta})}{\tau^\rho}\right)^2 d\tau\\
&=\int_0^1 \frac{1}{\tau^{2\rho+2}}\left(\frac{1}{2\beta}g^{(1)}(\frac{\log\tau}{-2\beta})+\rho g(\frac{\log\tau}{-2\beta})\right)^2 d\tau\\
%&=\int_0^\infty 2\beta e^{(4\rho+2)\beta t}\left(\frac{1}{2\beta}g^{(1)}(t)+\rho g(t)\right)^2 d\tau
\end{align*}
Taking the coordinate change $\tau = e^{-2\beta t}$ in the above equation and simple calculation yields (\ref{eq:dc_norm}). This completes the proof.

\subsection{Proof of Propositions \ref{prop:maxent_gsplinekernel-order1} and \ref{prop:maxent_dc}}

The proofs of Propositions \ref{prop:maxent_gsplinekernel-order1} and \ref{prop:maxent_dc} are similar. Due to the limitation of space, we only give the proof for the latter.

Define \begin{align}\label{eq:coordinatechange2}
l(t_i) = \frac{h(t_i)}{e^{-2\beta\rho t_i}}, i=0,\cdots,n-1
\end{align} Then  let $L=[l(t_0)\ l(t_1)\ \cdots\ l(t_{n-1})]^T$, $V=[h(t_0)\
h(t_1)\\\
\cdots\ h(t_{n-1})]^T$, and $B=\text{diag}([e^{-2\beta\rho t_0}\ e^{-2\beta\rho t_1}\ \cdots e^{-2\beta\rho t_{n-1}}])$. Then we have $V=BL$. Since $B$ is nonsingular, it holds that
$H(V)=H(L)+\log\det(B)$ according to
\cite[Corollary to Thm. 8.6.4]{CT12}). Moreover, since $B$ is independent of $l(t)$ and $h(t)$, the MaxEnt problem (\ref{eq:maxentdc}) is equivalent to
\begin{align*}
&\maximize_{l(\cdot)} \quad H(l(t_0),l(t_1),\cdots,l(t_{n-1})) \\
&\mbox{subject to}\quad \mean{l(t_i)}=0, i=0,\cdots,n-1, \\& \var{{l(t_{i+1})-{l(t_{i})}}}= e^{-2\beta t_i}-e^{-2\beta t_{i+1}},i=0,\cdots,n-2\\&
\var{l(t_{n-1})}=e^{-2\beta t_{n-1}},
\end{align*}
By \cite[Thm. 1]{CACCLP16}, the optimal solution to the above MaxEnt problem is the Gaussian process
\begin{equation}
\begin{aligned}\label{eq:sol2maxent4tc} l(t_k) &= \sum_{i=k}^{n-1} w(n-1-i) \sqrt{e^{-2\beta t_i} - e^{-2\beta t_{i+1}}},\\& k=0,\cdots,n-1,
l(t_n) =0 \text{ with } t_n=\infty.
\end{aligned}
\end{equation}
Clearly, (\ref{eq:sol2maxent4tc}) and (\ref{eq:coordinatechange2}) implies that  the Gaussian
process $g(t)$ in (\ref{eq:sol2maxent4dc}) is the optimal solution to the MaxEnt problem (\ref{eq:maxentdc}). This completes the proof.

\subsection{Proof of Proposition \ref{prop:bidiaginv}}

It follows from (\ref{eq:maxentdc}) that the Gaussian process $g(t)$ with $t\in\mathcal T$ defined in (\ref{eq:sol2maxent4dc}) can be rewritten as follows:
\begin{align}\nonumber
g&(t_{i+1}) = e^{-2\beta\rho(t_{i+1}-t_i)} g(t_{i}) \\\nonumber&\ +
e^{-2\beta\rho t_{i+1}}\sqrt{e^{-2\beta t_i}-e^{-2\beta t_{i+1}}} \bar w(i), i=0,\dots,n-2,\\g&(t_{n-1}) = e^{-\beta(2\rho+1) t_{n-1}} \bar w(n-1)\label{eq:DC_gp}\end{align} where $\bar w(i)$ with $i=0,1,\dots$ is a white Gaussian noise with zero mean and unit variance. It follows from (\ref{eq:DC_gp}) that the Gaussian process $g(t)$ with $t\in\mathcal T$ is also a Markov process with order 1. Then by \cite[Lem. A.3]{Chen16}, we have the kernel matrix $K^{\text{DC}}$ has tridiagonal inverse. This completes the proof.

\bibliographystyle{ieeetr}
\bibliography{ref}             % bib file to produce the bibliography
                                                     % with bibtex (preferred)

% that's all folks
\end{document}